\documentclass{article}

    \usepackage[final,nonatbib]{neurips_2022_ml4ps}

\usepackage[utf8]{inputenc} 
\usepackage[T1]{fontenc}    
\usepackage{hyperref}       
\usepackage{url}            
\usepackage{booktabs}       
\usepackage{amsmath, amsfonts, amssymb, mathrsfs, isomath, amsbsy, amssymb, amscd, amsfonts, dsfont, bm}      
\usepackage{nicefrac}       
\usepackage{microtype}      
\usepackage{xcolor}         

\usepackage{graphicx, relsize, subcaption, wrapfig}
\usepackage[style=ieee, backend=biber, citestyle=numeric-comp]{biblatex}
\title{Physics-informed neural networks for modeling rate- and temperature-dependent plasticity}

\author{%
    Rajat Arora$^{\dagger}$,$\quad$ Pratik Kakkar$^{\S}$,$\quad$ Amit Chakraborty$^{\S}$,$\quad$ Biswadip Dey$^{\S}$
    \\ \\
    $^{\dagger}$ Advanced Micro Devices (AMD), Austin, TX 78735, USA.
    \\
    \texttt{rajat.arora9464@gmail.com}
    \\ \\
    $^{\S}$ Siemens Technology, Princeton, NJ 08536, USA.
    \\
    \texttt{\{pratik.kakkar, amit.chakraborty, biswadip.dey\}@siemens.com}
}

\begin{document}

\maketitle

\begin{abstract}
This work presents a physics-informed neural network (PINN) based framework to model the strain-rate and temperature dependence of the deformation fields in elastic-viscoplastic solids. To avoid unbalanced back-propagated gradients during training, the proposed framework uses a simple strategy with no added computational complexity for selecting scalar weights that balance the interplay between different terms in the physics-based loss function. In addition, we highlight a fundamental challenge involving the selection of appropriate model outputs so that the mechanical problem can be faithfully solved using a PINN-based approach. We demonstrate the effectiveness of this approach by studying two test problems modeling the elastic-viscoplastic deformation in solids at different strain rates and temperatures, respectively. Our results show that the proposed PINN-based approach can accurately predict the spatio-temporal evolution of deformation in elastic-viscoplastic materials.
\end{abstract}
%
%

%
%
\section{Introduction}
%
Modeling the elastic-plastic response of materials using conventional numerical methods, such as finite element method, isogeometric analysis, or mesh-free methods, has always been computationally expensive due to the inherent iterative nature of discretization algorithms used in such methods. Furthermore, multitude of `fundamentally accurate' theories for the high-fidelity modeling of dislocation mediated plastic deformation at different scales \cite{nielsen2019finite, arora2020finite, fleck1994strain, arora2020unification, niordson2019homogenized, lynggaard2019finite,kuroda2008finite,evers2004non, arora2020dislocation,greer2013shear, arora2019computational, arora2022mechanics, joshi2020equilibrium} or fracture modeling in materials \cite{miehe2010thermodynamically,borden2014higher,yingjun2016phase,areias2016phase,kuhn2010continuum}, is bringing these numerical solvers to their limits. In this context, PINNs offer great opportunities to speed up (nonlinear) mechanical modeling of materials.

The idea of using neural networks to learn the solution of partial differential equations (PDEs) by minimizing a loss function, comprising the residual error of governing PDEs and its initial/boundary conditions, has been around for some time \cite{lagaris1998artificial, lagaris2000neural}. More recently, Raissi et.~al \cite{raissi2017physicsI, raissi2019physics} have extended this concept towards PINNs which can solve the forward and inverse problems involving general nonlinear PDEs by relying on small or even zero labeled datasets. Several applications of PINNs can be found in the literature ranging from modeling of fluid flows and Navier Stokes equations \cite{sun2020surrogate,rao2020physics,zhang2020frequencycompensated,jin2021nsfnets,gao2020phygeonet,zhang2020frequency}, cardiovascular systems \cite{kissas2020machine,sahli2020physics}, and material modeling \cite{frankel_prediction_2020, arora2022physrnet,  tipireddy2019comparative,zhang2020physics,meng2020composite,zhu2021machine, arora2021machine}, among others. Compared to traditional data-driven approaches for predicting path-dependent plastic behavior in metals \cite{mozaffar2019deep, deep_learning_plasticity2, Lstm_plasticity2, Huang_pod_cmame, gorji_potention_plast}, PINNs can learn high-fidelity surrogate models while simultaneously reducing (or even eliminating) the need for bigger training datasets. However, developing a physics-informed neural network to model the spatio-temporal variation of deformation in elastic-plastic solids, along with its dependence on strain-rate and temperature, poses several technical challenges.

In this work, we take a first step in highlighting these challenges and demonstrate the strength of PINNs for modeling elastic-viscoplastic deformation in materials. In particular, we focus on predicting the spatio-temporally varying deformation fields (displacement, stress, and plastic strain) under different strain rates (i.e., applied loading rate) and temperatures, respectively. We present a detailed discussion on the construction of (physics-based) composite loss along with a brief summary on ways to avoid unbalanced back-propagated (exploding) gradients during model training. Furthermore, a strategy with no added computational complexity for choosing the scalar weights that balance the interplay between different terms in composite loss is also proposed. Although the current work focuses on the scenarios with monotonic loading paths, we note that the deformation of an elastic-viscoplastic solid is a highly nonlinear function of temperature, strain rate, spatial coordinates, and strain. This \textit{real-time} stress predictive capability for elastic-viscoplastic materials enjoys special use in the design and development of energy storage devices (e.g., lithium metal solid-state-batteries). Specifically, the study conducted here corresponds to analyzing the effect of impact (i.e.~crash) and heat to the solid lithium anode in the solid state batteries.

\noindent \textbf{\textit{Notation and Terminology}:}
Vectors and tensors are represented by bold face lower- and upper-case letters, respectively. The symbol `$\cdot$' denotes single contraction of adjacent indices of two tensors (i.e. $\boldsymbol{a} \cdot \boldsymbol{b} = a_i b_i$ or $\boldsymbol{A} \cdot \boldsymbol{n} = A_{ij}n_j$). The symbol `$:$' denotes double contraction of adjacent indices of two tensors of rank two or higher (i.e. $\boldsymbol{A}:\boldsymbol{B} = A_{ij}B_{ij}$ or $\bm{\mathbb{C}}:\boldsymbol{A} = \mathbb{C}_{ijkl}A_{kl}$). The norm of a second order tensor $\boldsymbol{A}$ is given by $||\boldsymbol{A}|| = \sqrt{\boldsymbol{A}:\boldsymbol{A}}$. The symbols $\boldsymbol{\nabla}$ and $Div$ denote the gradient and the divergence operators, respectively. $\boldsymbol{I}$ denotes the second order identity tensor.
%
%

%
%
\section{Background on the Deformation Behavior of Elastic-viscoplastic Solids}
\label{sec:equations}
%
In this section, we describe the nonlinear PDEs that govern the behavior of elastic-viscoplastic solids under loads at small deformation (see \cite{gurtin_fried_anand_2010} for further details). In the absence of body and inertial forces, the strong form of the mechanical equilibrium on a volumetric domain $\Omega$ can be expressed as
\begin{equation}
\begin{aligned}
Div\,\boldsymbol{\sigma} &= \bm{0} ~~\text{in}~~ \Omega, ~~\text{with}
\\
\boldsymbol{\sigma}\cdot\boldsymbol{n} &= \boldsymbol{t}_{bc} ~~ \text{on}~~\partial\Omega_{N} \text{~~and~~}	\boldsymbol{u} = \boldsymbol{u}_{bc}~~ \text{on}~~\partial\Omega_{D},
\label{eq:equilibrium}
\end{aligned}
\end{equation}
where $\boldsymbol{\sigma}$ and $\boldsymbol{u}$ denote the stress and displacement, respectively, $\boldsymbol{n}$ denote the unit outward normal to the external boundary $\partial\Omega$ of domain $\Omega$, and $\boldsymbol{t}_{bc}$ and $\boldsymbol{u}_{bc}$ denote the known traction and displacement vectors on the Neumann boundary $\partial\Omega_{N}$ and Dirichlet boundary $\partial\Omega_{D}$, respectively. The total strain tensor $\boldsymbol{\epsilon}$, which is the symmetric part of the displacement gradient, can be decomposed into the sum of elastic and plastic strain components denoted by $\boldsymbol{\epsilon}^e$ and $\boldsymbol{\epsilon}^p$ , respectively, i.e., $\boldsymbol{\nabla}\boldsymbol{u} + (\boldsymbol{\nabla}\boldsymbol{u})^T = 2(\boldsymbol{\epsilon}^e + \boldsymbol{\epsilon}^p)$. The stress is given by the Hooke's law $
\boldsymbol{\sigma} = \mathbb{C}:\boldsymbol{\epsilon}^e$, where $\mathbb{C}$ is the fourth order elasticity tensor. Furthermore, the plastic strain evolution is governed by 
\begin{equation}
\dot{\boldsymbol{\epsilon}}^p 
= 
\left(\frac{3}{2}\right)^{\frac{m+1}{2m}}\,A e^{ -\frac{Q}{R\theta}}\frac{\boldsymbol{\sigma}'}{||\boldsymbol{\sigma}'||}\bigg({\frac{||\boldsymbol{\sigma}'||}{S}}\bigg)^{1/m},
\label{eq:eqp_evol}
\end{equation}
where $\boldsymbol{\sigma}' = \boldsymbol{\sigma} - \text{trace}(\boldsymbol{\sigma})\boldsymbol{I}$ denotes the deviatoric part of the stress tensor, $A$ is a pre-exponential factor, $Q$ denotes the activation energy, $R$ is the molar gas constant, $\theta$ denotes the temperature of the domain $\Omega$, and $m \in (0, 1]$ is a strain-rate-sensitivity parameter.

Then, by letting $S$ denote the material strength, its dynamics can be expressed as
\begin{equation}
\dot{S} = h(S, \boldsymbol{\sigma}),
\label{eq:s_evol}
\end{equation}
with the hardening function $h(S, \boldsymbol{\sigma})$ defined as \cite{anand2019elastic},  
\begin{align*}
h({S, \boldsymbol{\sigma}}) 
= 
\left[ H_0 \left| 1- \frac{S}{S_s}\right|^a \text{sign}\left(1- \frac{S}{S_s}\right) \right]	\dot{\bar{\epsilon}}^p,
\end{align*}
where $H_0$ and $n$ are two strain-hardening parameters. Furthermore, the saturation value of $S$ for a given strain rate and temperature, i.e., $S_s$, is given by $S_s = S_* (e^{\frac{Q}{R\theta}} \dot{\bar{\epsilon}}^p /A)^n$, where $S_*$ and $a$ are two additional strain-hardening parameters.

In this work, we select the material parameters as: $A = 4.25\times10^4$ s$^{-1}$, $Q=37$ kJ/mol, $m=0.15$, $S_*=2$ MPa, $S_0=0.95$ MPa, $H_0=10$ MPa, $a=2$, $n=0.05$, $E=7810$ MPa, and $\nu=0.38$. These parameters have been calibrated using the experimental data from direct tension tests on polycrystalline lithium specimens \cite{anand2019elastic, lepage2019lithium}.
%
%

%
%
\section{Proposed Learning Framework}
\label{sec:architecture}
%
Since plastic strain $\boldsymbol{\epsilon}^p$ and stress $\boldsymbol{\sigma}$ are related by Hooke's law, the elastic-plastic deformation can be uniquely characterized by the displacement vector $\boldsymbol{u}$ and any one of the two tensors, $\boldsymbol{\epsilon}^p$ or $\boldsymbol{\sigma}$, along with the internal variable $S$. However, as shown in Appendix~\ref{app:A}, a PINN with such choice of outputs suffers from degraded accuracy and convergence issues. Therefore, we \textit{propose a mixed-variable formulation} and use $\boldsymbol{u}$, $\boldsymbol{\sigma}$, $\boldsymbol{\epsilon}^p$ and $S$ as the PINN outputs. To model the effect of strain rate and temperature on the elastic-viscoplastic behavior in two-dimensional solids, we use two separate PINNs to predict the output variables at any given location $(x_1, x_2)$. To capture strain rate dependence, the first PINN uses scalar strain $\Gamma$ and strain rate $\hat{\Gamma}$ as two more additional inputs. The second PINN uses scalar strain $\Gamma$ and temperature $\theta$ as the additional inputs to capture temperature dependence. These PINNs are realized via a multilayer perceptron with 9 hidden layers, 120 neurons/layer, and $tanh$ activation function. In addition, we normalize the data along each component.

\begin{wrapfigure}[43]{r}{0.7\textwidth}
\vspace{-1.5em}
	\centering
	\begin{subfigure}[b]{.14\linewidth}
		\centering
		\tiny \text{$\Gamma = 0.01$}\par
		{\includegraphics[width=0.995\linewidth]{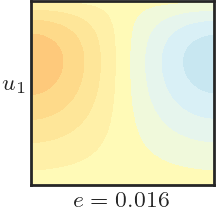}}
	\end{subfigure}
	\begin{subfigure}[b]{.14\linewidth}
		\centering
		\tiny \text{$\Gamma = 0.02$}\par
		{\includegraphics[width=0.995\linewidth]{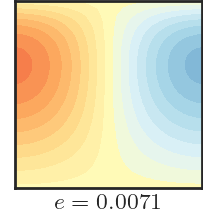}}
	\end{subfigure}
	\begin{subfigure}[b]{.14\linewidth}
		\centering
		\tiny \text{$\Gamma = 0.04$}\par
		{\includegraphics[width=0.995\linewidth]{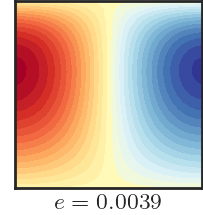}}
	\end{subfigure}
	~~~~
	\begin{subfigure}[b]{.14\linewidth}
		\centering
		\tiny \text{$\Gamma = 0.01$}\par
		{\includegraphics[width=0.995\linewidth]{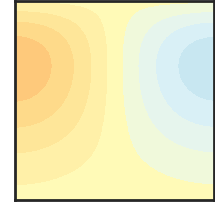}}
	\end{subfigure}
	\begin{subfigure}[b]{.14\linewidth}
		\centering
		\tiny \text{$\Gamma = 0.02$}\par
		{\includegraphics[width=0.995\linewidth]{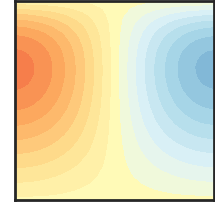}}
	\end{subfigure}
	\begin{subfigure}[b]{.14\linewidth}
		\centering
		\tiny \text{$\Gamma = 0.04$}\par
		{\includegraphics[width=0.995\linewidth]{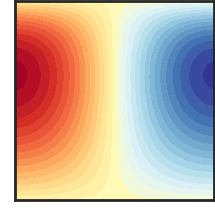}}
	\end{subfigure}\\[.02em]

	\centering
	\begin{subfigure}[b]{.14\linewidth}
		\centering
		{\includegraphics[width=0.995\linewidth]{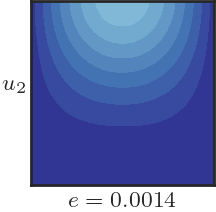}}
	\end{subfigure}
	\begin{subfigure}[b]{.14\linewidth}
		\centering
		{\includegraphics[width=0.995\linewidth]{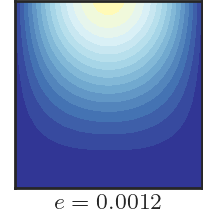}}
	\end{subfigure}
	\begin{subfigure}[b]{.14\linewidth}
		\centering
		{\includegraphics[width=0.995\linewidth]{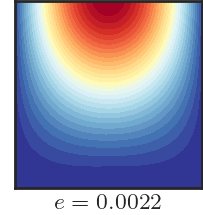}}
	\end{subfigure}~~~~
	\begin{subfigure}[b]{.14\linewidth}
		\centering
		{\includegraphics[width=0.995\linewidth]{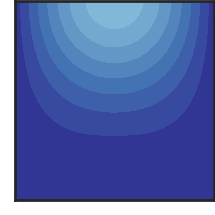}}
	\end{subfigure}
	\begin{subfigure}[b]{.14\linewidth}
		\centering
		{\includegraphics[width=0.995\linewidth]{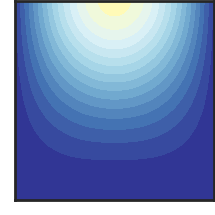}}
	\end{subfigure}
	\begin{subfigure}[b]{.14\linewidth}
		\centering
		{\includegraphics[width=0.995\linewidth]{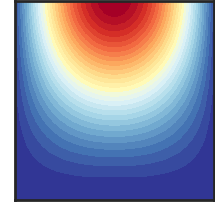}}
	\end{subfigure}\\[.02em]

	\centering
	\begin{subfigure}[b]{.14\linewidth}
		\centering
		{\includegraphics[width=0.995\linewidth]{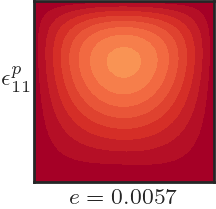}}
	\end{subfigure}
	\begin{subfigure}[b]{.14\linewidth}
		\centering
		{\includegraphics[width=0.995\linewidth]{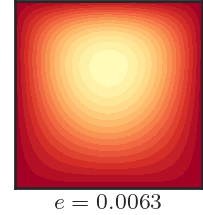}}
	\end{subfigure}
	\begin{subfigure}[b]{.14\linewidth}
		\centering
		{\includegraphics[width=0.995\linewidth]{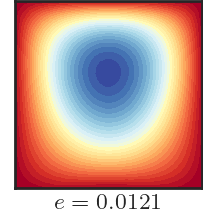}}
	\end{subfigure}~~~~
	\begin{subfigure}[b]{.14\linewidth}
		\centering
		{\includegraphics[width=0.995\linewidth]{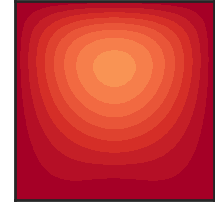}}
	\end{subfigure}
	\begin{subfigure}[b]{.14\linewidth}
		\centering
		{\includegraphics[width=0.995\linewidth]{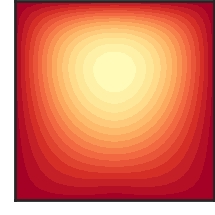}}
	\end{subfigure}
	\begin{subfigure}[b]{.14\linewidth}
		\centering
		{\includegraphics[width=0.995\linewidth]{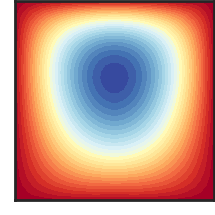}}
	\end{subfigure}\\[.02em]

	\centering
	\begin{subfigure}[b]{.14\linewidth}
		\centering
		{\includegraphics[width=0.995\linewidth]{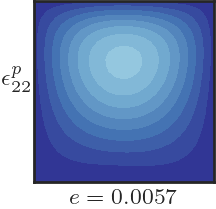}}
	\end{subfigure}
	\begin{subfigure}[b]{.14\linewidth}
		\centering
		{\includegraphics[width=0.995\linewidth]{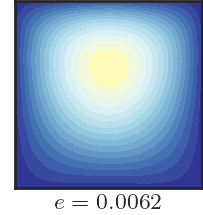}}
	\end{subfigure}
	\begin{subfigure}[b]{.14\linewidth}
		\centering
		{\includegraphics[width=0.995\linewidth]{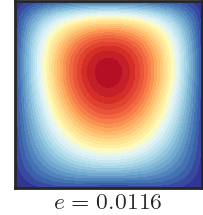}}
	\end{subfigure}~~~~
	\begin{subfigure}[b]{.14\linewidth}
		\centering
		{\includegraphics[width=0.995\linewidth]{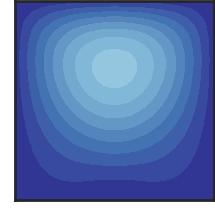}}
	\end{subfigure}
	\begin{subfigure}[b]{.14\linewidth}
		\centering
		{\includegraphics[width=0.995\linewidth]{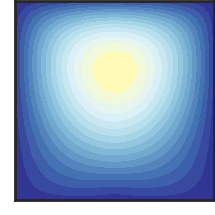}}
	\end{subfigure}
	\begin{subfigure}[b]{.14\linewidth}
		\centering
		{\includegraphics[width=0.995\linewidth]{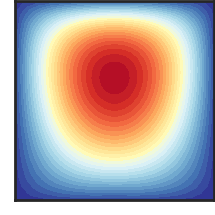}}
	\end{subfigure}\\[.02em]

	\centering
	\begin{subfigure}[b]{.14\linewidth}
		\centering
		{\includegraphics[width=0.995\linewidth]{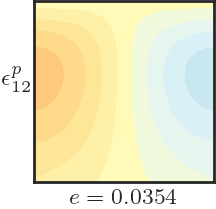}}
	\end{subfigure}
	\begin{subfigure}[b]{.14\linewidth}
		\centering
		{\includegraphics[width=0.995\linewidth]{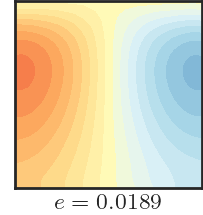}}
	\end{subfigure}
	\begin{subfigure}[b]{.14\linewidth}
		\centering
		{\includegraphics[width=0.995\linewidth]{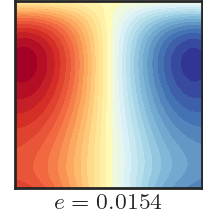}}
	\end{subfigure}~~~~
	\begin{subfigure}[b]{.14\linewidth}
		\centering
		{\includegraphics[width=0.995\linewidth]{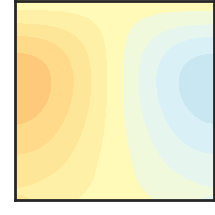}}
	\end{subfigure}
	\begin{subfigure}[b]{.14\linewidth}
		\centering
		{\includegraphics[width=0.995\linewidth]{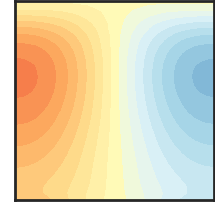}}
	\end{subfigure}
	\begin{subfigure}[b]{.14\linewidth}
		\centering
		{\includegraphics[width=0.995\linewidth]{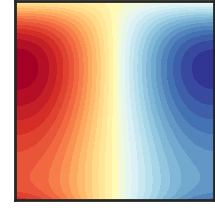}}
	\end{subfigure}\\[.02em]

	\centering
	\begin{subfigure}[b]{.14\linewidth}
		\centering
		{\includegraphics[width=0.995\linewidth]{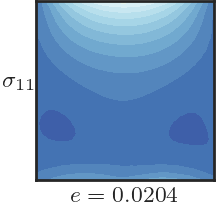}}
	\end{subfigure}
	\begin{subfigure}[b]{.14\linewidth}
		\centering
		{\includegraphics[width=0.995\linewidth]{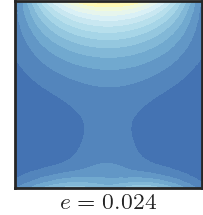}}
	\end{subfigure}
	\begin{subfigure}[b]{.14\linewidth}
		\centering
		{\includegraphics[width=0.995\linewidth]{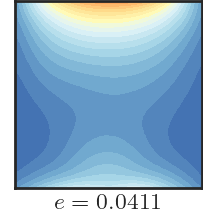}}
	\end{subfigure}~~~~
	\begin{subfigure}[b]{.14\linewidth}
		\centering
		{\includegraphics[width=0.995\linewidth]{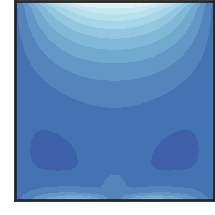}}
	\end{subfigure}
	\begin{subfigure}[b]{.14\linewidth}
		\centering
		{\includegraphics[width=0.995\linewidth]{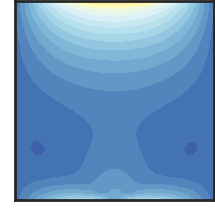}}
	\end{subfigure}
	\begin{subfigure}[b]{.14\linewidth}
		\centering
		{\includegraphics[width=0.995\linewidth]{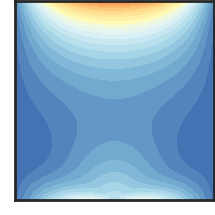}}
	\end{subfigure}\\[.02em]

	\centering
	\begin{subfigure}[b]{.14\linewidth}
		\centering
		{\includegraphics[width=0.995\linewidth]{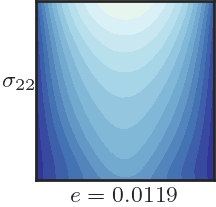}}
	\end{subfigure}
	\begin{subfigure}[b]{.14\linewidth}
		\centering
		{\includegraphics[width=0.995\linewidth]{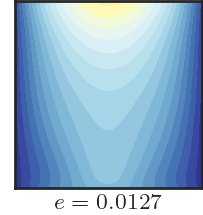}}
	\end{subfigure}
	\begin{subfigure}[b]{.14\linewidth}
		\centering
		{\includegraphics[width=0.995\linewidth]{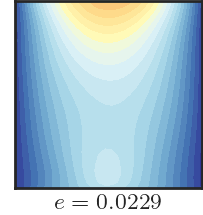}}
	\end{subfigure}~~~~
	\begin{subfigure}[b]{.14\linewidth}
		\centering
		{\includegraphics[width=0.995\linewidth]{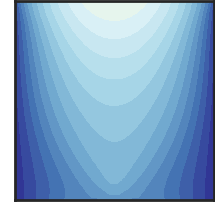}}
	\end{subfigure}
	\begin{subfigure}[b]{.14\linewidth}
		\centering
		{\includegraphics[width=0.995\linewidth]{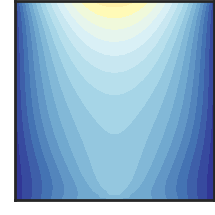}}
	\end{subfigure}
	\begin{subfigure}[b]{.14\linewidth}
		\centering
		{\includegraphics[width=0.995\linewidth]{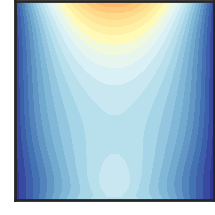}}
	\end{subfigure}\\[.02em]

	
	\centering
	\begin{subfigure}[b]{.14\linewidth}
		\centering
		{\includegraphics[width=0.995\linewidth]{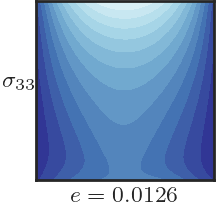}}
	\end{subfigure}
	\begin{subfigure}[b]{.14\linewidth}
		\centering
		{\includegraphics[width=0.995\linewidth]{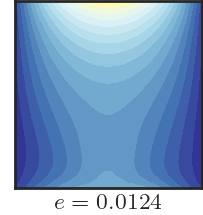}}
	\end{subfigure}
	\begin{subfigure}[b]{.14\linewidth}
		\centering
		{\includegraphics[width=0.995\linewidth]{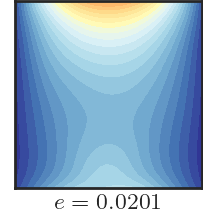}}
	\end{subfigure}~~~~
	\begin{subfigure}[b]{.14\linewidth}
		\centering
		{\includegraphics[width=0.995\linewidth]{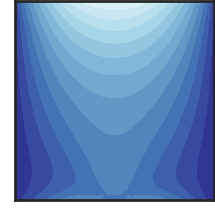}}
	\end{subfigure}
	\begin{subfigure}[b]{.14\linewidth}
		\centering
		{\includegraphics[width=0.995\linewidth]{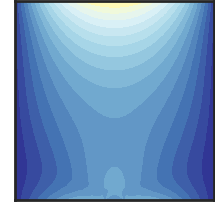}}
	\end{subfigure}
	\begin{subfigure}[b]{.14\linewidth}
		\centering
		{\includegraphics[width=0.995\linewidth]{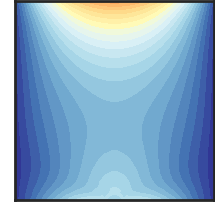}}
	\end{subfigure}\\[.02em]

	\centering
	\begin{subfigure}[b]{.14\linewidth}
		\centering
		{\includegraphics[width=0.995\linewidth]{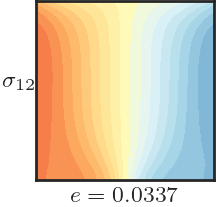}}
	\end{subfigure}
	\begin{subfigure}[b]{.14\linewidth}
		\centering
		{\includegraphics[width=0.995\linewidth]{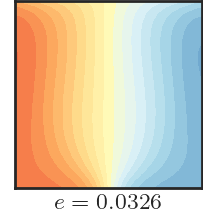}}
	\end{subfigure}
	\begin{subfigure}[b]{.14\linewidth}
		\centering
		{\includegraphics[width=0.995\linewidth]{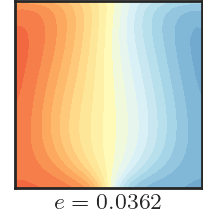}}
	\end{subfigure}~~~~
	\begin{subfigure}[b]{.14\linewidth}
		\centering
		{\includegraphics[width=0.995\linewidth]{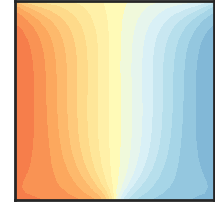}}
	\end{subfigure}
	\begin{subfigure}[b]{.14\linewidth}
		\centering
		{\includegraphics[width=0.995\linewidth]{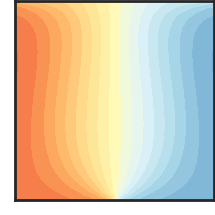}}
	\end{subfigure}
	\begin{subfigure}[b]{.14\linewidth}
		\centering
		{\includegraphics[width=0.995\linewidth]{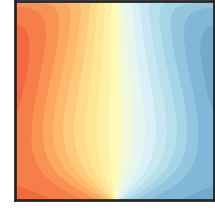}}
	\end{subfigure}\\[.02em]

	\centering
	\begin{subfigure}[b]{.14\linewidth}
		\centering
		{\includegraphics[width=0.995\linewidth]{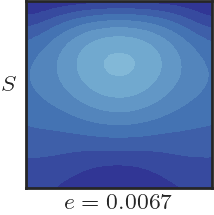}}
	\end{subfigure}
	\begin{subfigure}[b]{.14\linewidth}
		\centering
		{\includegraphics[width=0.995\linewidth]{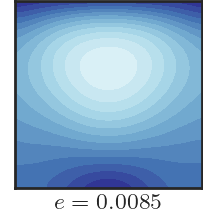}}
	\end{subfigure}
	\begin{subfigure}[b]{.14\linewidth}
		\centering
		{\includegraphics[width=0.995\linewidth]{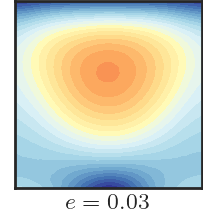}}
	\end{subfigure}~~~~
	\begin{subfigure}[b]{.14\linewidth}
		\centering
		{\includegraphics[width=0.995\linewidth]{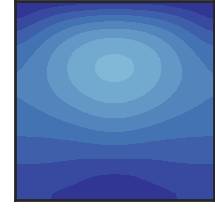}}
	\end{subfigure}
	\begin{subfigure}[b]{.14\linewidth}
		\centering
		{\includegraphics[width=0.995\linewidth]{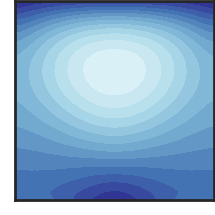}}
	\end{subfigure}
	\begin{subfigure}[b]{.14\linewidth}
		\centering
		{\includegraphics[width=0.995\linewidth]{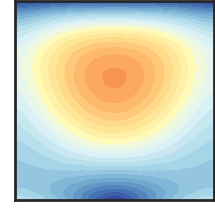}}
	\end{subfigure}\\[.04em]
	
	\begin{subfigure}[b]{.8\linewidth}
		\centering
		{\includegraphics[width=0.85\linewidth]{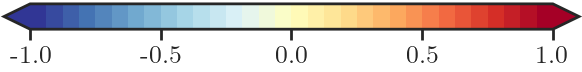}}
	\end{subfigure}
	\vspace{-0.5em}
        \caption{\small{Comparison of the results obtained from the PINN model (left block) with the ground truth reference data (right block) for $\hat\Gamma = 10^{-2.5}\, s^{-1}$.}}
	\label{fig:res_strain_rate}
	\vspace{-2em}
\end{wrapfigure}

\textbf{Physics-informed Loss Function:} The composite loss function is defined as the weighted sum of a physics loss $\mathcal{L}_{phy}$ and the data loss computed as the mean-squared-error (MSE) between the normalized ground truth data and PINN outputs. The physics loss is composed of seven different components - \\
i) $\mathcal{L}_{pde}$: PDE loss, \\
ii) $\mathcal{L}_{Dbc}$: Dirichlet boundary condition loss,\\
iii) $\mathcal{L}_{Nbc}$:  Neumann boundary condition loss,\\
iv) $\mathcal{L}_{ic}$: initial condition loss,\\
v) $\mathcal{L}_{c}$: constitutive loss corresponding to the satisfaction of constitutive law,\\
vi) $\mathcal{L}_{psr}$: plastic strain rate loss corresponding to the equation governing the evolution of plastic strain, and \\
vii) $\mathcal{L}_{s}$:  strength loss enforcing the material strength evolution equation. \\
Please see Appendix~\ref{sec:construction_loss} for further details on the construction of the loss function.

\textbf{Network Training:} After initializing the network with Xavier initialization \cite{glorot2010understanding}, we train it on PyTorch using Adam optimizer \cite{kingma_adam:_2015} (\textit{initial learning rate} = $10^{-3}$). The network is trained for $8000$ epochs during which the learning rate is varied using \texttt{ReduceLROnPlateau} scheduler (\textit{patience}=$30$). 

\textbf{Dataset Generation:} To generate the ground truth data, we develop an in-house code using deal.II \cite{BangerthHartmannKanschat2007} and solve equations \eqref{eq:equilibrium}-\eqref{eq:s_evol} over a $32\times 32$ grid for scalar strain values up to $\Gamma = 0.04$. Furthermore, for studying the effect of strain rate and temperature on the spatio-temporal evolution of deformation fields in the body, the ground truth dataset is generated for strain rates $\hat\Gamma$ ($\text{s}^{-1}$): $\{10^{-4}, 10^{-3}, 10^{-2}, 10^{-1}\}$ at $\theta = 298K$ and for temperatures $\theta$ ($K$): $\{298, 318, 358, 378\}$ at $\hat\Gamma = 10^{-1} \text{s}^{-1}$. The dataset is then randomly split into a $80:20$ ratio for training and validation purposes.
\begin{wrapfigure}[44]{r}{0.7\textwidth}
\vspace{-.75em}
	\centering
	\begin{subfigure}[b]{.14\linewidth}
		\centering
		\tiny \text{$\Gamma = 0.01$}\par
		{\includegraphics[width=0.995\linewidth]{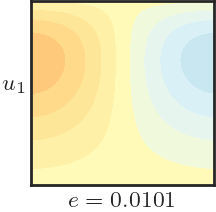}}
	\end{subfigure}
	\begin{subfigure}[b]{.14\linewidth}
		\centering
		\tiny \text{$\Gamma = 0.02$}\par
		{\includegraphics[width=0.995\linewidth]{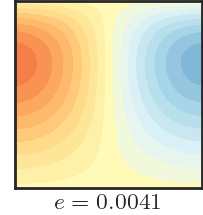}}
	\end{subfigure}
	\begin{subfigure}[b]{.14\linewidth}
		\centering
		\tiny \text{$\Gamma = 0.04$}\par
		{\includegraphics[width=0.995\linewidth]{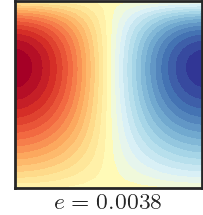}}
	\end{subfigure}~~~~
	\begin{subfigure}[b]{.14\linewidth}
		\centering
		\tiny \text{$\Gamma = 0.01$}\par
		{\includegraphics[width=0.995\linewidth]{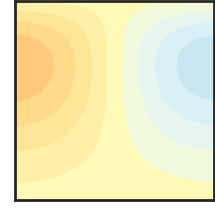}}
	\end{subfigure}
	\begin{subfigure}[b]{.14\linewidth}
		\centering
		\tiny \text{$\Gamma = 0.02$}\par
		{\includegraphics[width=0.995\linewidth]{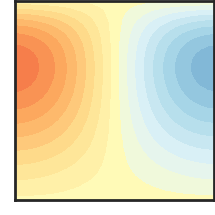}}
	\end{subfigure}
	\begin{subfigure}[b]{.14\linewidth}
		\centering
		\tiny \text{$\Gamma = 0.04$}\par
		{\includegraphics[width=0.995\linewidth]{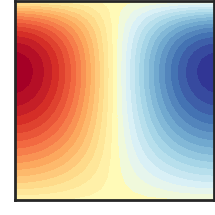}}
	\end{subfigure}\\[.02em]

	\centering
	\begin{subfigure}[b]{.14\linewidth}
		\centering
		{\includegraphics[width=0.995\linewidth]{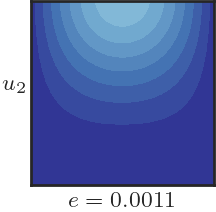}}
	\end{subfigure}
	\begin{subfigure}[b]{.14\linewidth}
		\centering
		{\includegraphics[width=0.995\linewidth]{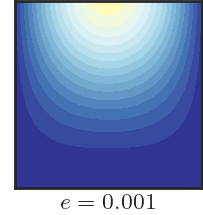}}
	\end{subfigure}
	\begin{subfigure}[b]{.14\linewidth}
		\centering
		{\includegraphics[width=0.995\linewidth]{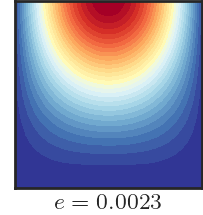}}
	\end{subfigure}~~~~
	\begin{subfigure}[b]{.14\linewidth}
		\centering
		{\includegraphics[width=0.995\linewidth]{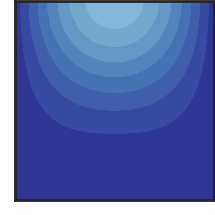}}
	\end{subfigure}
	\begin{subfigure}[b]{.14\linewidth}
		\centering
		{\includegraphics[width=0.995\linewidth]{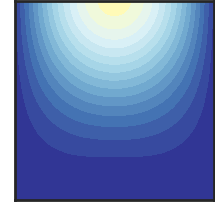}}
	\end{subfigure}
	\begin{subfigure}[b]{.14\linewidth}
		\centering
		{\includegraphics[width=0.995\linewidth]{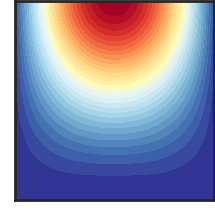}}
	\end{subfigure}\\[.02em]

	\centering
	\begin{subfigure}[b]{.14\linewidth}
		\centering
		{\includegraphics[width=0.995\linewidth]{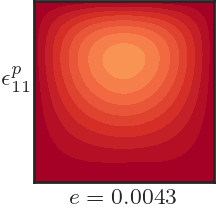}}
	\end{subfigure}
	\begin{subfigure}[b]{.14\linewidth}
		\centering
		{\includegraphics[width=0.995\linewidth]{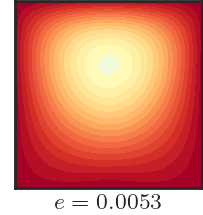}}
	\end{subfigure}
	\begin{subfigure}[b]{.14\linewidth}
		\centering
		{\includegraphics[width=0.995\linewidth]{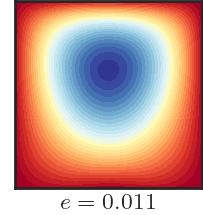}}
	\end{subfigure}~~~~
	\begin{subfigure}[b]{.14\linewidth}
		\centering
		{\includegraphics[width=0.995\linewidth]{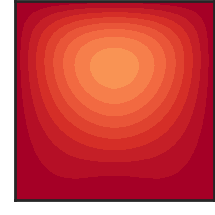}}
	\end{subfigure}
	\begin{subfigure}[b]{.14\linewidth}
		\centering
		{\includegraphics[width=0.995\linewidth]{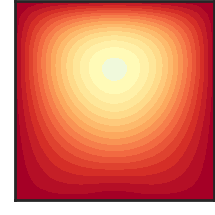}}
	\end{subfigure}
	\begin{subfigure}[b]{.14\linewidth}
		\centering
		{\includegraphics[width=0.995\linewidth]{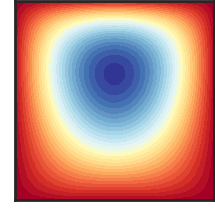}}
	\end{subfigure}\\[.02em]

	\centering
	\begin{subfigure}[b]{.14\linewidth}
		\centering
		{\includegraphics[width=0.995\linewidth]{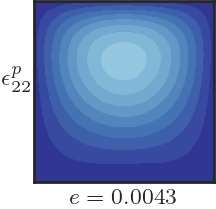}}
	\end{subfigure}
	\begin{subfigure}[b]{.14\linewidth}
		\centering
		{\includegraphics[width=0.995\linewidth]{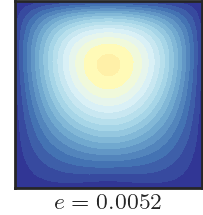}}
	\end{subfigure}
	\begin{subfigure}[b]{.14\linewidth}
		\centering
		{\includegraphics[width=0.995\linewidth]{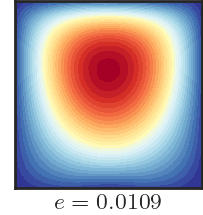}}
	\end{subfigure}~~~~
	\begin{subfigure}[b]{.14\linewidth}
		\centering
		{\includegraphics[width=0.995\linewidth]{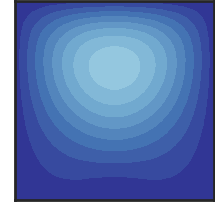}}
	\end{subfigure}
	\begin{subfigure}[b]{.14\linewidth}
		\centering
		{\includegraphics[width=0.995\linewidth]{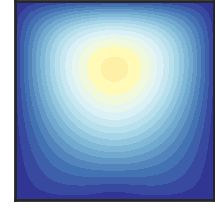}}
	\end{subfigure}
	\begin{subfigure}[b]{.14\linewidth}
		\centering
		{\includegraphics[width=0.995\linewidth]{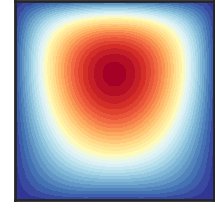}}
	\end{subfigure}\\[.02em]

	\centering
	\begin{subfigure}[b]{.14\linewidth}
		\centering
		{\includegraphics[width=0.995\linewidth]{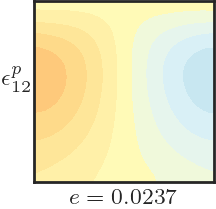}}
	\end{subfigure}
	\begin{subfigure}[b]{.14\linewidth}
		\centering
		{\includegraphics[width=0.995\linewidth]{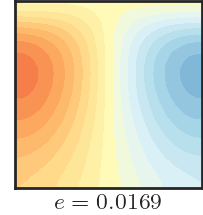}}
	\end{subfigure}
	\begin{subfigure}[b]{.14\linewidth}
		\centering
		{\includegraphics[width=0.995\linewidth]{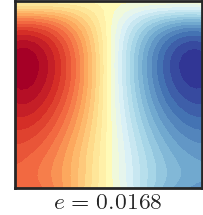}}
	\end{subfigure}~~~~
	\begin{subfigure}[b]{.14\linewidth}
		\centering
		{\includegraphics[width=0.995\linewidth]{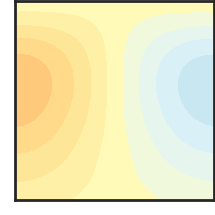}}
	\end{subfigure}
	\begin{subfigure}[b]{.14\linewidth}
		\centering
		{\includegraphics[width=0.995\linewidth]{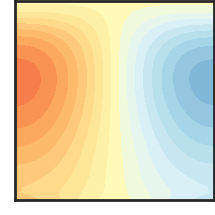}}
	\end{subfigure}
	\begin{subfigure}[b]{.14\linewidth}
		\centering
		{\includegraphics[width=0.995\linewidth]{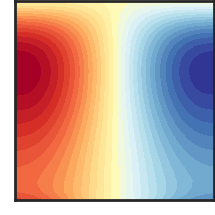}}
	\end{subfigure}\\[.02em]

	\centering
	\begin{subfigure}[b]{.14\linewidth}
		\centering
		{\includegraphics[width=0.995\linewidth]{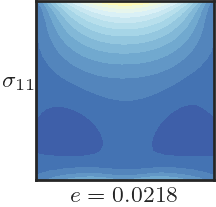}}
	\end{subfigure}
	\begin{subfigure}[b]{.14\linewidth}
		\centering
		{\includegraphics[width=0.995\linewidth]{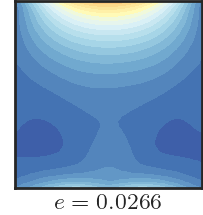}}
	\end{subfigure}
	\begin{subfigure}[b]{.14\linewidth}
		\centering
		{\includegraphics[width=0.995\linewidth]{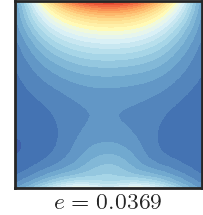}}
	\end{subfigure}~~~~
	\begin{subfigure}[b]{.14\linewidth}
		\centering
		{\includegraphics[width=0.995\linewidth]{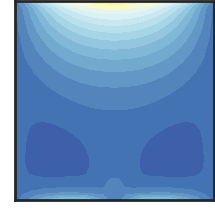}}
	\end{subfigure}
	\begin{subfigure}[b]{.14\linewidth}
		\centering
		{\includegraphics[width=0.995\linewidth]{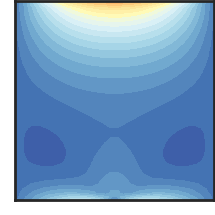}}
	\end{subfigure}
	\begin{subfigure}[b]{.14\linewidth}
		\centering
		{\includegraphics[width=0.995\linewidth]{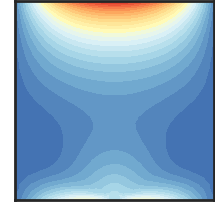}}
	\end{subfigure}\\[.02em]

	\centering
	\begin{subfigure}[b]{.14\linewidth}
		\centering
		{\includegraphics[width=0.995\linewidth]{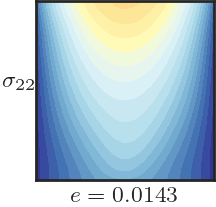}}
	\end{subfigure}
	\begin{subfigure}[b]{.14\linewidth}
		\centering
		{\includegraphics[width=0.995\linewidth]{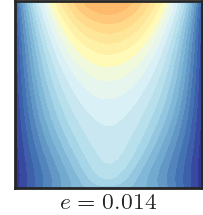}}
	\end{subfigure}
	\begin{subfigure}[b]{.14\linewidth}
		\centering
		{\includegraphics[width=0.995\linewidth]{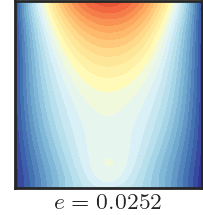}}
	\end{subfigure}~~~~
	\begin{subfigure}[b]{.14\linewidth}
		\centering
		{\includegraphics[width=0.995\linewidth]{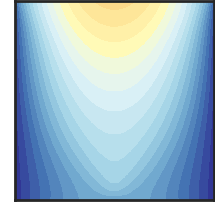}}
	\end{subfigure}
	\begin{subfigure}[b]{.14\linewidth}
		\centering
		{\includegraphics[width=0.995\linewidth]{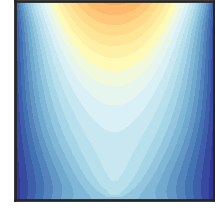}}
	\end{subfigure}
	\begin{subfigure}[b]{.14\linewidth}
		\centering
		{\includegraphics[width=0.995\linewidth]{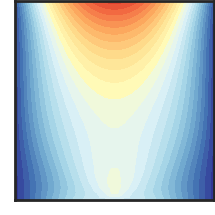}}
	\end{subfigure}\\[.02em]

	\centering
	\begin{subfigure}[b]{.14\linewidth}
		\centering
		{\includegraphics[width=0.995\linewidth]{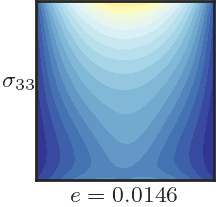}}
	\end{subfigure}
	\begin{subfigure}[b]{.14\linewidth}
		\centering
		{\includegraphics[width=0.995\linewidth]{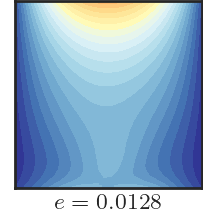}}
	\end{subfigure}
	\begin{subfigure}[b]{.14\linewidth}
		\centering
		{\includegraphics[width=0.995\linewidth]{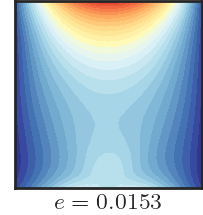}}
	\end{subfigure}~~~~
	\begin{subfigure}[b]{.14\linewidth}
		\centering
		{\includegraphics[width=0.995\linewidth]{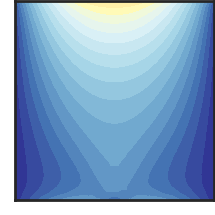}}
	\end{subfigure}
	\begin{subfigure}[b]{.14\linewidth}
		\centering
		{\includegraphics[width=0.995\linewidth]{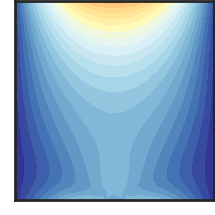}}
	\end{subfigure}
	\begin{subfigure}[b]{.14\linewidth}
		\centering
		{\includegraphics[width=0.995\linewidth]{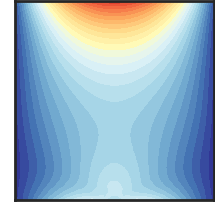}}
	\end{subfigure}\\[.02em]

	\centering
	\begin{subfigure}[b]{.14\linewidth}
		\centering
		{\includegraphics[width=0.995\linewidth]{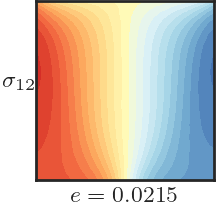}}
	\end{subfigure}
	\begin{subfigure}[b]{.14\linewidth}
		\centering
		{\includegraphics[width=0.995\linewidth]{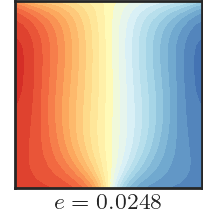}}
	\end{subfigure}
	\begin{subfigure}[b]{.14\linewidth}
		\centering
		{\includegraphics[width=0.995\linewidth]{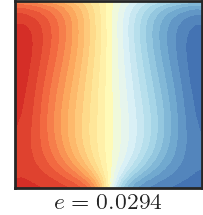}}
	\end{subfigure}~~~~
	\begin{subfigure}[b]{.14\linewidth}
		\centering
		{\includegraphics[width=0.995\linewidth]{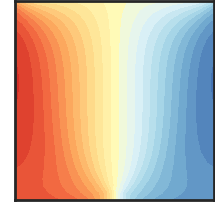}}
	\end{subfigure}
	\begin{subfigure}[b]{.14\linewidth}
		\centering
		{\includegraphics[width=0.995\linewidth]{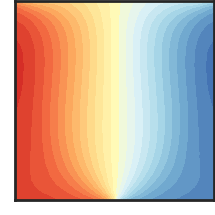}}
	\end{subfigure}
	\begin{subfigure}[b]{.14\linewidth}
		\centering
		{\includegraphics[width=0.995\linewidth]{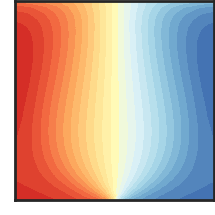}}
	\end{subfigure}\\[.02em]

	\centering
	\begin{subfigure}[b]{.14\linewidth}
		\centering
		{\includegraphics[width=0.995\linewidth]{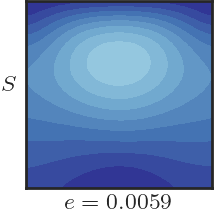}}
	\end{subfigure}
	\begin{subfigure}[b]{.14\linewidth}
		\centering
		{\includegraphics[width=0.995\linewidth]{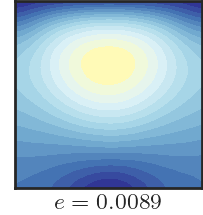}}
	\end{subfigure}
	\begin{subfigure}[b]{.14\linewidth}
		\centering
		{\includegraphics[width=0.995\linewidth]{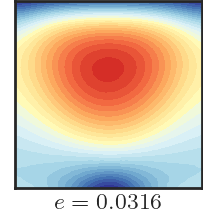}}
	\end{subfigure}~~~~
	\begin{subfigure}[b]{.14\linewidth}
		\centering
		{\includegraphics[width=0.995\linewidth]{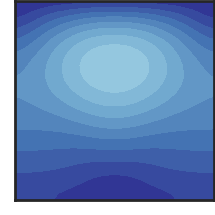}}
	\end{subfigure}
	\begin{subfigure}[b]{.14\linewidth}
		\centering
		{\includegraphics[width=0.995\linewidth]{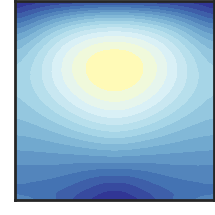}}
	\end{subfigure}
	\begin{subfigure}[b]{.14\linewidth}
		\centering
		{\includegraphics[width=0.995\linewidth]{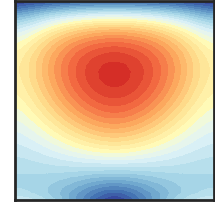}}
	\end{subfigure}\\[.02em]
	
	\begin{subfigure}[b]{.8\linewidth}
		\centering
		{\includegraphics[width=0.85\linewidth]{./figures/color_bar.png}}
	\end{subfigure}\\[.02em]
	
	\vspace{-0.5em}
	\caption{\small{Comparison of the results obtained from the PINN model (left block) and the ground truth reference data (right block) for $T = 328 K$.}}
	\label{fig:res_temperature}
	\vspace{-2em}
\end{wrapfigure}
%
%

%
%
%
\section{Results \& Discussion}
\label{sec:results}
%
%
%
\textbf{Case I: Strain rate dependence} First we compare the predicted values of the stress, plastic strain, and displacement fields in the domain with a test dataset for $\hat\Gamma=10^{-2.5} s^{-1}$ at $\Gamma = 0.01, 0.02,$ and $0.04$ (Figure~\ref{fig:res_strain_rate}). We can notice that the predicted values have no visible artifacts and are in great agreement with the FEM reference results. Small values of the normalized root-mean-squared-error $e$ (reported underneath the corresponding field plot) further confirms predictions by the PINN match the FEM reference results remarkably well.

\begin{wrapfigure}[10]{r}{0.45\textwidth}
\vspace{-1em}
\centering
\includegraphics[width=0.9\linewidth]{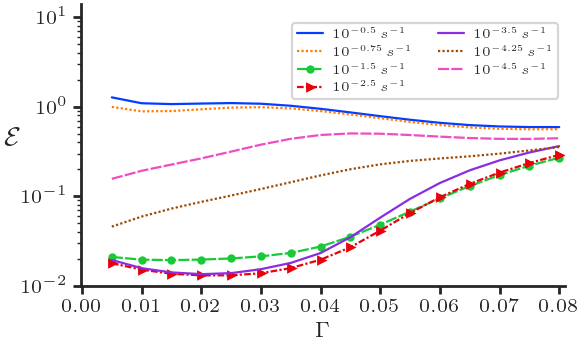}
\vspace{-1em}
\captionof{figure}{\small{Variation of error $\mathcal{E}$ with strain $\Gamma$ for different strain rates in and outside of the training range ($\hat\Gamma \in [10^{-4}, 10^{-1}] s^{-1}$, $\Gamma \in [0, 0.04]$).}}
\label{fig:error_strain_rate}
\vspace{-2em}
\end{wrapfigure}

Next, we test the predictive power of the trained PINN for values of inputs that lie outside the training data range. Specifically, we calculate the averaged value of the normalized root-mean-squared-error $\mathcal{E}$ for $7$ different strain rate values $\hat\Gamma (s^{-1}) = \{10^{-0.5}, \allowbreak 10^{-.75}, \allowbreak10^{-1.5}, \allowbreak 10^{-2.5}, \allowbreak 10^{-3.5}, 10^{-4.25}, 10^{-4.5}\}$ and multiple strain values in the range $[0, 0.08]$. We can readily notice from Figure~\ref{fig:error_strain_rate} that the error $\mathcal{E}$ is very small $(\approx 1\%)$ upto strain $\Gamma = 0.04$ when strain rate lies within the training range. However, in the region $\Gamma\in(0.04, 0.08]$, the error $\mathcal{E}$ steadily increases to $\approx 10\%$. Also, when the strain rate is outside the training data range, the error $\mathcal{E}$ is large at all strains implying that the predicted values do not match well with the FEM data.
%
%

%
%
\textbf{Case II: Temperature dependence} Similar to \textit{Case I}, we first compare the predicted values of the stress, plastic strain, and displacement fields in the domain with a test dataset for $T=328 K$ at $\Gamma = 0.01, 0.02,$ and $0.04$ (Figure~\ref{fig:res_temperature}). We can notice that the error $e$ has small values and the predictions match well with the FEM reference results.

\begin{wrapfigure}[14]{r}{0.45\textwidth}
\centering
\includegraphics[width=0.85\linewidth]{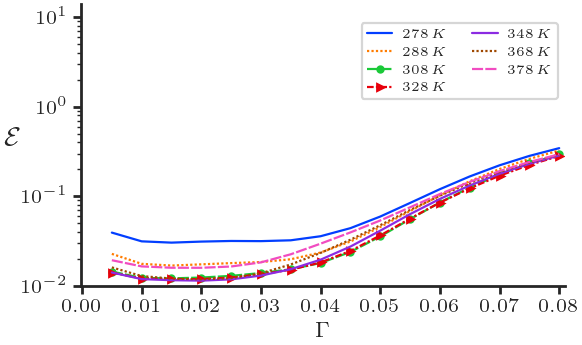}
\label{-3em}
\captionof{figure}{\small{Variation of error $\mathcal{E}$ with strain $\Gamma$ for different temperatures in and outside of the training data range ($\theta \in [298, 378] K$, $\Gamma \in [0, 0.04]$).}}
\label{fig:error_temperature}
\vspace{-1em}
\end{wrapfigure}

Figure \ref{fig:error_temperature} shows that the error $\mathcal{E}$ rises to $\ge 10\%$ as the strains go beyond the training data range. On the other hand, when the strain is within the training range, the errors are still $< 10\%$ even when the temperature is outside the training range. 
%
%
%

%
%
\section{Conclusion}
\label{sec:conclusion}
%
%
This work demonstrates the strength of PINNs in problems dealing with the evolution of highly nonlinear deformation field in elastic-viscoplastic materials under monotonous loading. In particular, we trained two specific PINN models and applied them predicting the spatio-temporally varying deformation field in elastic-viscoplastic materials at different strain rates, and temperatures, respectively. The predicted values are in great agreement with the ground truth reference data for the test cases discussed in this work.

This work highlighted a fundamental challenge involving selection of appropriate model outputs so that the mechanical problem can be faithfully solved using neural networks. We present and compare two potential choice of outputs for the model in Appendix \ref{app:A} and present detailed reasoning for preferring one choice over the other. This work also discusses the construction of composite loss function, comprising the data loss component and physics-based loss components. We also use a novel physics-based strategy for selecting the non-dimensional scalar constants that weigh each component in the physics-based loss function without any added computational complexity. Moreover, a novel loss criterion for residual calculation corresponding to plastic strain rate equation is proposed to alleviate issues related to unbalanced back-propagated (exploding) gradients during model training.

The real-time stress field prediction in such highly nonlinear mechanical system paves the way for many new applications, such as design and optimization of lithium ion batteries or inverse modeling problems which were previously computationally intractable.
%
%

%
%
\section*{Broader Impact}
We introduce a learning framework for predicting nonlinear spatio-temporal variation of deformation inside elastic-viscoplastic materials under various operating conditions (e.g., temperature, strain rate, etc.). This is an important problem in multiple engineering applications. For example, the outcome of this work can be used to carry out fast (near real-time) simulation of the behavior of a solid $Li$ anode under varying loading conditions. This, in turn, can improve the process of design, development, and operation of solid-state lithium-metal batteries.\ However, our proposed framework is still a conceptual proposal and has a very low (around 2) Technology Readiness Level (TRL) \cite{hirshorn2016final}. We are yet to fully understand its limitations and failure scenarios that can significantly influence its real-world adoption.

{
\small
\printbibliography
}

%
%
%
%
\appendix
%
%

%
%
\section*{Appendix}
\section{Construction of the Physics-informed Loss Function}
\label{sec:construction_loss}
%
The development of a PINN based approach to predict the solution of a system of nonlinear PDEs can be viewed as an optimization problem which involves solving for $(\boldsymbol{W}, \boldsymbol{b})$ that minimizes network's total loss. The composite loss $\mathcal{L}$ comprises the summation of supervised data loss $\mathcal{L}_{data}$ and the physics-based loss $\mathcal{L}_{phy}$ i.e.~$\mathcal{L} = \mathcal{L}_{data} + \mathcal{L}_{phy}$. The non-dimensional supervised data loss $\mathcal{L}_{data}$ measures the discrepancy between the normalized ground truth data $\bar{\boldsymbol{T}}$ and the neural network outputs $\bar{\boldsymbol{Y}}$ and is given by 
\begin{equation}
\mathcal{L}_{data} = \frac{1}{N_{data}} \sum^{N_{data}}_{j=1} \sum^{N_{out}}_{i = 1} || \bar Y^{(j)}_i - \bar{T}^{(j)}_i ||^2,
\end{equation}
where $N_{data}$ denotes the number of ground truth samples and $N_{out}$ is the number of scalar output variables.

To evaluate the physics-based loss $\mathcal{L}_{phy}$, we sample a collection of randomly distributed collocation points discretizing the normalized input space. The whole set of collocation points is denoted by $\mathcal{P} = \{\mathcal{P}_\omega, \mathcal{P}_{\zeta, D}, \mathcal{P}_{\zeta, N}, \mathcal{P}_{\eta}\}$ where $\mathcal{P}_\omega$ denotes the collocation points in the entire input space $[-1, 1]^{N_{in}}$. $\mathcal{P}_{\zeta, D}$ and $\mathcal{P}_{\zeta, N}$ denote the subset of $\mathcal{P}$ that intersects with the $\partial\Omega_D$ and $\partial\Omega_N$, respectively. $\mathcal{P}_{\eta}$ denotes the subset of $\mathcal{P}$ that intersects with $\Gamma = 0 ~(\text{or}\,t = 0)$. 

To this end, we construct the physics-based loss $\mathcal{L}_{phy}$ with seven components i) PDE loss $\mathcal{L}_{pde}$, ii) Dirichlet boundary condition loss $\mathcal{L}_{Dbc}$, iii) Neumann boundary condition loss $\mathcal{L}_{Nbc}$, iv) initial condition loss $\mathcal{L}_{ic}$, v) constitutive loss $\mathcal{L}_{c}$ corresponding to the satisfaction of constitutive law, vi) plastic strain rate loss $\mathcal{L}_{psr}$ corresponding to the equation governing the evolution of plastic strain, and vii) strength loss $\mathcal{L}_{s}$ enforcing the material strength evolution equation. Each component of $\mathcal{L}_{phy}$ is individually calculated as follows:
\begin{align}
\mathcal{L}_{pde} 
&=
|| Div\,\boldsymbol{\sigma} ||^2_{\mathcal{P}_\omega}
\nonumber \\
\mathcal{L}_{Dbc} 
&= 
|| \boldsymbol{u} - \boldsymbol{u}_{bc}||^2_{\mathcal{P}_{\zeta,D}}
\nonumber \\
\mathcal{L}_{Nbc} 
&= 
|| \boldsymbol{\sigma} \cdot\boldsymbol{n} - \boldsymbol{t}_{bc} ||^2_{\mathcal{P}_{\zeta,N}}
\nonumber \\
\mathcal{L}_{ic} 
&=
|| \boldsymbol{Y} - \mathcal{I}_0||^2_{\mathcal{P}_\eta}
\label{eq:phy_loss}
\\
\mathcal{L}_{c} 
&= 
|| \boldsymbol{\sigma} - \mathbb{C} : (\boldsymbol{\nabla}\boldsymbol{u} - \boldsymbol{\epsilon}^p) ||^2
\nonumber \\
\mathcal{L}_{s} 
&= 
|| \dot{S} - h(S, \boldsymbol{\sigma})||^2
\nonumber \\
\mathcal{L}_{psr} 
&=
\mathrm{MMSE}\left( 
\mathsmaller{
\dot{\boldsymbol{\epsilon}}_p - \sqrt{\frac{3}{2}} \frac{A\boldsymbol{\sigma}'}{||\boldsymbol{\sigma}'||} e^{-\frac{Q}{R\theta}}\left({\sqrt\frac{3}{2}\frac{||\boldsymbol{\sigma}'||}{S}}\right)^{1/m}
}
\right)
\nonumber 
\end{align}    
In the above, $\mathcal{I}_0$ denotes the initial state of the system i.e.\,outputs at $t=0$. The loss criterion $\mathrm{MMSE}$ is discussed in detail below. $\mathcal{L}_{phy}$ is then given as the weighted sum of these loss components
\begin{equation}
\mathcal{L}_{phy} 
= 
\lambda_1 \mathcal{L}_{pde} + \lambda_2\mathcal{L}_{ic} + \lambda_3 \mathcal{L}_{psr} + \lambda_4\mathcal{L}_{s}
+ \lambda_5\mathcal{L}_c + \lambda_6 \mathcal{L}_{Dbc} + \lambda_7 \mathcal{L}_{Nbc}
\end{equation}
where $\lambda_i,\; (i = 1...6)$ are the scalar weights. Please note that each of these loss components are computed using automatic differentiation.

Next, we briefly discuss the two main difficulties that hinder the training of DNNs for elastic-viscoplastic modeling applications.\\
\textbf{(I)} The power law dependence of the equivalent plastic strain rate $\dot\epsilon^p$ leads to large values of $L^2$ norm of $\mathcal{L}_{psr}$ loss $(\ge O(10^{18}))$ which causes unstable imbalance in the magnitude of the back-propagated gradients during the training when using common loss criterions such as Mean-Squared-Error $(\mathrm{MSE})$. Therefore, in this work, we use a novel Modified Mean Squared Error (MMSE) loss criterion to reduce the numerical stiffness associated with equation \eqref{eq:eqp_evol} and allow stable gradients to be used during the training
\begin{equation}
	\text{MMSE}(A) = log_{10}(1 + ||A||).
\end{equation}
In the above, $A$ denotes the residual value. The loss criterion is equivalent to the Mean Squared Error (MSE) criterion when the discrepancy between the residual values are small. 
\\
\textbf{(II)} The relative coefficients $\lambda_i$ $(i = 1..6)$ for all the losses comprising $\mathcal{L}_{phy}$ play an important role in mitigating the gradient pathology issue during the training \cite{wang2020understanding}. There are competing effects between these different loss components which can lead to convergence issues during the minimization of the composite loss $\mathcal{L}$ (see \cite[Sec.~4.1]{sun2020surrogate}). 	While the recent advances in mitigating gradient pathologies \cite{wang2020understanding, bischof2021multi} might improve predictive accuracy, they introduce additional computational and memory overhead because of the calculation of an adaptive factor for each loss component. In this work, we devise a simple strategy, with no added computational complexity, to evaluate the coefficients which remain constant during the course of training. The strategy is outlined as follows:
\begin{itemize}
	\item The Dirichlet boundary condition and initial condition losses ($\mathcal{L}_{ic}$ and $\mathcal{L}_{Dbc}$) are calculated in a normalized manner (scaled between $[-1,1]$). So, we take $\lambda_2 = \lambda_6 = 1$.
	\item The other loss components are nondimensionalized using appropriate scales as shown in Table \ref{tab:nondim}. $\mu_c$ is a constant chosen to scale quantities with units of stress. Based on the observation that stress is often nondimensionalized by Shear Modulus $\mu$ in conventional numerical methods, we choose $\mu_c = 0.01 \mu$ to achieve tight tolerance on the equilibrium equation and traction boundary conditions. 
	\item Since material strength $S$ and $\mu$ differ by orders of magnitude, $S$ is nondimensionalized by $S_0$.
	\item We nondimensionalize time by using strain rate $\hat\Gamma$, since $\hat\Gamma$ sets the time scale for the problem.
	\item The length is nondimensionalized by the characteristic length of the domain, chosen to be $H$ in this work.
\end{itemize}
\begin{table}[h]
\begin{center}
\begin{tabular}{ c c}
\hline\\[-.9em]
	Loss component &Scaling \\ 		\hline\\[-.8em]
	$\mathcal{L}_{pde}$ & $\lambda_1 = \frac{H^2}{{\mu_c}^2}$ \\[.4em] \hline \\[-.8em]
	$\mathcal{L}_{ic}$ & $\lambda_2 = 1$ \\[.4em] \hline \\[-.8em] 
	$\mathcal{L}_{psr}$ & $\lambda_3 = \frac{1}{{\hat\Gamma}^2}$ \\[.4em] \hline \\[-.8em] 
	$\mathcal{L}_{s}$ & $\lambda_4 = \frac{1}{{(S_0\,\hat\Gamma)}^2}$ \\[.5em] \hline \\[-.8em]
	$\mathcal{L}_{c}$ & $\lambda_5 = \frac{1}{{\mu_c}^2}$ \\[.4em] \hline \\[-.8em]
	$\mathcal{L}_{Dbc}$ & $\lambda_6 = 1$ \\[.4em] \hline \\[-.8em] 
	$\mathcal{L}_{Nbc}$ & $\lambda_7 = \frac{1}{{\mu_c}^2}$ \\[.4em] \hline \\[-.8em]
\end{tabular}
\end{center}
\caption{Scaling constants for different physics-based loss components.}
\label{tab:nondim}
\vspace{-2em}
\end{table}
%
%

%
%
\section{The Rationale Governing the Selection of Output Variables}
\label{app:A}
%
This section compares the results obtained from two PINN models - (a) Model I with displacement, stress, plastic strain, and strength  ($\boldsymbol{u}, \boldsymbol{\sigma}, \boldsymbol{\epsilon}^p, S$) as outputs; and (b) Model II with displacement, plastic strain, and stress ($\boldsymbol{u}, \boldsymbol{\epsilon}^p, S$) as outputs. 

For Model II, the physics-based loss $\mathcal{L}_{phy}$ is obtained from the set of equations \eqref{eq:phy_loss} with the following important changes: i) The stress is directly calculated from the displacements and plastic strains which are outputs of the neural network,  i.e.~$\boldsymbol{\sigma}=\mathbb{C}:(\boldsymbol{\nabla}\boldsymbol{u} - \boldsymbol{\epsilon}^p)$. This implicitly leads to satisfaction of constitutive law so the loss component $\mathcal{L}_c$ is ignored. ii) The data loss $\mathcal{L}_{data}$ is  also modified to account for the current model outputs.  

The study conducted here corresponds to Case I. i.e., understanding the effect of strain rate on the spatio-temporal evolution of deformation in an elastic-viscoplastic material. The learning rate for Model II is taken to be $10^{-4}$ while keeping the collocation points and all other hyperparameters the same for both the architectures as described in Section~\ref{sec:results}. 
\begin{figure}[h!]
\centering
{\includegraphics[width=.6\linewidth]{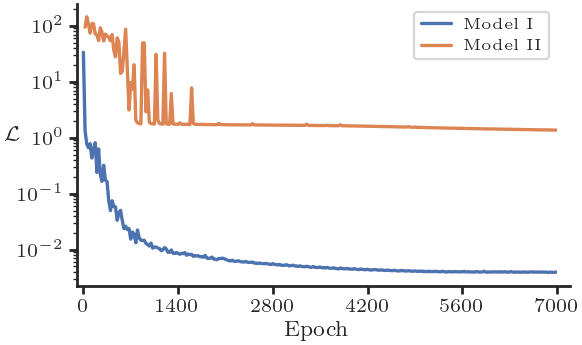}}
\caption{\small{Comparison of training history for Model I and II differing only in the model outputs.}}
\label{fig:loss_comp_app}
\vspace{-1.75em}
\end{figure}

The convergence of the training loss for both the models is presented in Fig.~\ref{fig:loss_comp_app}. It can be seen that loss reaches a stagnation value of $\approx\!1.9$ for model II at around $3000$ epochs which is approximately hundred times larger than the converged loss value obtained for model I.  We can conclude that model I  does not suffer from any such degraded accuracy or convergence issue as indicated by Figure \ref{fig:loss_comp_app}. This result is an extension of the similar observation for the purely linear elastic calculations presented in \cite{rao2021physics} to the general elastic-plastic modeling case discussed here.
	
While the exact reasons for such a behavior are still unclear, we highlight the main differences between the two models. First, the stress calculated in model II is sensitive to the noise in the gradients of $\boldsymbol{u}$.  Second, we note that highest order of the spatial derivatives occurring in the composite loss function is one and two for models I and II, respectively. Moreover, in elastic/elastic-plastic deformations the order of displacement field magnitudes in the $x_1$ and $x_2$ direction can be vastly different because of the loading setup and Poisson's effect.  We believe that these factors combine together to give rise to convergence issue and degraded accuracy when using model II.  The use of improved training technique \cite{czarnecki2017sobolev}, which also approximates target derivatives along with target values, may alleviate these issues for model II but that may involve added computational complexity and remains the subject of future investigation.
%
%

%
%
\section{Ablation Studies on the Effect of Network Depth and Width}
\label{app:AblationNetSize}
%
First, we perform a (non-exhaustive) parametric study to identify a suitable number of hidden layers $N_l$ and number of neurons per layer $N_n$ needed to model the deformation field with an acceptable
\begin{figure}[h!]
\centering
\includegraphics[width=0.6\linewidth]{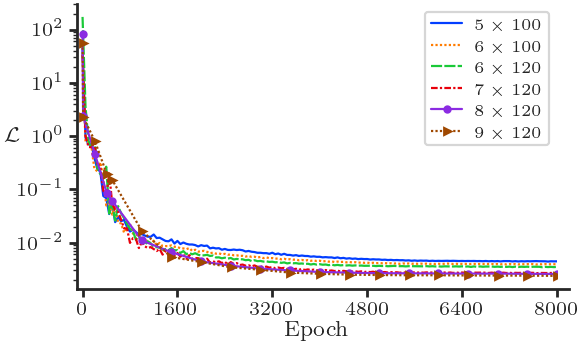}
\vspace{-1em}
\captionof{figure}{\small{Comparison of the composite loss $\mathcal{L}$ associated with different PINN architectures $(N_l \times N_n)$ for Case I. $N_l$ and $N_n$ denotes the number of layers and number of nodels per layer, repectively.}}
\label{fig:arch_strain_rate}
\vspace{-1em}
\end{figure}
accuracy. We train $6$ neural networks with the following architectures: i) $5\times 100$, ii) $6 \times 100$, iii) $6\times 120$, iv) $7\times 120$, v) $8\times 120$, and vi) $9\times 120$. Figure \ref{fig:arch_strain_rate} presents the training history for each of these architectures. As expected, we see a merit in increasing both $N_n$ and $N_l$ initially but the final value of the composite loss stops improving when the number of layers are increased from $7$ to $9$ keeping $N_n$ fixed at $120$. These three network architectures ($7\times120, 8\times120, $ and $9\times120$) reduce the nondimensional composite loss by almost five orders of magnitude (from $10^2$ to $\sim\!10^{-3}$). The values of the corresponding validation losses are monitored to notice any overfitting issues. We use the neural network with architecture $9\times120$ for generating results associated with Case I in Section~\ref{sec:results}.

\begin{figure}[h!]
\centering
\includegraphics[width=0.6\linewidth]{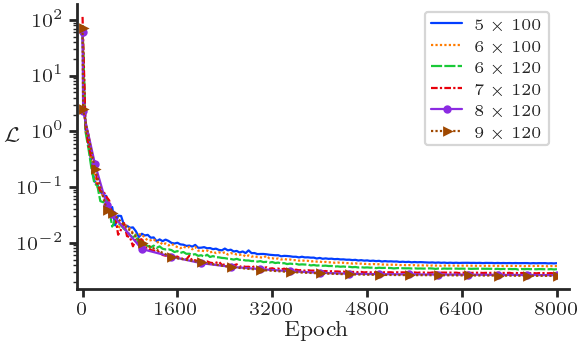}
\captionof{figure}{\small{Comparison of the composite loss $\mathcal{L}$ associated with different PINN architectures $(N_l \times N_n)$ for Case II.}}
\label{fig:arch_temperature}
\vspace{-1.5em}
\end{figure}

Similar to Case I, we first conduct a study to gain insight into the effect of $N_l$ and $N_n$ on the composite loss $\mathcal{L}$ and train the six aforementioned neural network architectures (i.e., $5\times 100$, $6 \times 100$, $6\times 120$, $7\times 120$, $8\times 120$, $9\times 120$). Figure \ref{fig:arch_temperature} presents the training history for each of these architectures which shows similar trend as in Figure \ref{fig:arch_strain_rate}. Therefore, we use the neural network with architecture $9\times120$ for generating results associated with Case II in Section~\ref{sec:results}.

\newpage
\section*{Checklist}

%
\begin{enumerate}

\item For all authors...
\begin{enumerate}
  \item Do the main claims made in the abstract and introduction accurately reflect the paper's contributions and scope?
    \answerYes{}
  \item Did you describe the limitations of your work?
    \answerYes{}
  \item Did you discuss any potential negative societal impacts of your work?
    \answerNA{}
  \item Have you read the ethics review guidelines and ensured that your paper conforms to them?
    \answerYes{}
\end{enumerate}

\item If you are including theoretical results...
\begin{enumerate}
  \item Did you state the full set of assumptions of all theoretical results?
    \answerNA{}
        \item Did you include complete proofs of all theoretical results?
    \answerNA{}
\end{enumerate}

\item If you ran experiments...
\begin{enumerate}
  \item Did you include the code, data, and instructions needed to reproduce the main experimental results (either in the supplemental material or as a URL)?
    \answerNo{}
  \item Did you specify all the training details (e.g., data splits, hyperparameters, how they were chosen)?
    \answerYes{See section 2 and 3.}
        \item Did you report error bars (e.g., with respect to the random seed after running experiments multiple times)?
    \answerNo{}
        \item Did you include the total amount of compute and the type of resources used (e.g., type of GPUs, internal cluster, or cloud provider)?
    \answerNo{}
\end{enumerate}

\item If you are using existing assets (e.g., code, data, models) or curating/releasing new assets...
\begin{enumerate}
  \item If your work uses existing assets, did you cite the creators?
    \answerYes{}
  \item Did you mention the license of the assets?
    \answerNo{}
  \item Did you include any new assets either in the supplemental material or as a URL?
    \answerNo{}
  \item Did you discuss whether and how consent was obtained from people whose data you're using/curating?
    \answerNA{}
  \item Did you discuss whether the data you are using/curating contains personally identifiable information or offensive content?
    \answerNA{}
\end{enumerate}

\item If you used crowdsourcing or conducted research with human subjects...
\begin{enumerate}
  \item Did you include the full text of instructions given to participants and screenshots, if applicable?
    \answerNA{}
  \item Did you describe any potential participant risks, with links to Institutional Review Board (IRB) approvals, if applicable?
    \answerNA{}
  \item Did you include the estimated hourly wage paid to participants and the total amount spent on participant compensation?
    \answerNA{}
\end{enumerate}

\end{enumerate}

\end{document}